\documentclass[runningheads]{llncs}

\usepackage{booktabs} 

\usepackage{amsfonts}
\usepackage{color}
\usepackage{graphicx}
\usepackage{amsmath}

\newcommand{\hr}{{\hat r}}

\begin{document}
\title{Conformative Filtering for Implicit Feedback Data}
\author{Farhan Khawar \and
Nevin L. Zhang}
\authorrunning{F. Khawar, N. L. Zhang}
\institute{Department of Computer Science and Engineering\\ The Hong Kong University of Science and Technology, Hong Kong \\
\email{\{fkhawar,lzhang\}@cse.ust.hk}}
\maketitle

\begin{abstract}
Implicit feedback is the simplest form of user feedback that can be used for item recommendation. It is easy to collect and is domain independent. However, there is a lack of negative examples. Previous work tackles this problem by assuming that users are not interested or not as much interested in the unconsumed items. Those assumptions are often severely violated since non-consumption can be due to factors like unawareness or lack of resources. Therefore, non-consumption by a user does not always mean disinterest or irrelevance. In this paper, we propose a novel method called Conformative Filtering  (CoF) to address the issue. The motivating observation is that if there is a large group of users who share the same taste and none of them have consumed an item before, then it is likely that the item is not of interest to the group. We perform multidimensional clustering on implicit feedback data using hierarchical latent tree analysis (HLTA) to identify user ``taste'' groups and make recommendations for a user based on her memberships in the groups and on the past behavior of the groups. Experiments on two real-world datasets from different domains show that CoF has superior performance compared to several common baselines.
\end{abstract}


	\section{Introduction}
	
	With the advent of the online marketplace, an average user is presented with an un-ending choice of items to consume.
	Those could be  books to buy, web-pages to click, songs to listen,  movies to watch, and so on. Online stores and content providers no longer have to worry about shelf space to display their items.  However, too much choice is not always a luxury. It can also be an unwanted distraction and makes it difficult for a user to find the items she desires. It is necessary to automatically filter a vast amount of items and identify those that are of interest to a user.

	Collaborative filtering (CF) \cite{Goldberg:1992:UCF:138859.138867}	
	 is one commonly used technique to deal with the problem. Most research work on CF focuses on explicit feedback data, where ratings on items have been previously provided by users \cite{Koren2015}.
	  Items with high ratings are preferred over those with low ratings. In other words,  items with high ratings are positive examples, while those with low ratings are negative examples. Unrated items are  missing data.

	In practice, one often encounters implicit feedback data, where users did not explicitly rate items \cite{nichols1997implicit}. Recommendations need to be made based on user activities such as clicks, page views, and purchase actions. Those are  positive-only data and contain information regarding which items were consumed. There is no information about the unconsumed items. In other words, there are no negative examples. The problem is hence called {\em one class collaborative filtering (OCCF)} \cite{Pan:2008:OCF:1510528.1511402}.

In previous works, the lack of negative examples in OCCF is addressed by adopting one of the following four strategies with respect to each user: (1) Treat unconsumed items as negative examples \cite{ning2011slim}; (2) Treat unconsumed items as negative examples with low confidence \cite{Hu:2008:CFI:1510528.1511352}; (3) Identify some unconsumed items as negative examples using heuristics \cite{Pan:2008:OCF:1510528.1511402}; (4) Assume the user prefers consumed items over unconsumed items \cite{Rendle:2009:BBP:1795114.1795167}. We refer to the strategies  as the {\em unconsumed as negative (UAN)}, {\em  UAN-with-low-confidence}, {\em UAN-with-chance} and
{\em consumed preferred over unconsumed (CPU)} assumptions respectively.

All the assumptions are problematic. The UAN assumption is in
contradiction with the very objective of collaborative filtering --- to identify items that might
be of interest to a user among those she did not consume before. Moreover, if we assume a user does not like two items to exactly the same degree, then theoretically there is 50\% chance that she would prefer the next item she chooses to consume to the last item she consumed.

 In this paper, we adopt a new assumption:
 If there is a large group of users who share the same taste and
none of them  have consumed an item before, then  the item is not of interest to the group.
  By a {\em taste} we mean the tendency to
	consume a certain collection of items such as comedy movies, pop songs, or spicy food.
We call our assumption the {\em group UAN} assumption because it is with respect to a user group. In contrast,  we refer to the first assumption mentioned above  as the {\em individual UAN} because it is with respect to an individual user. Group UAN is more reasonable than individual UAN because there is less chance of treating unawareness as disinterest.

We identify user taste groups by performing multidimensional clustering using
hierarchical latent class analysis (HLTA) \cite{chen2017latent}. HLTA can detect sets of items that tend to be {\em co-consumed} in the sense users who consumed some of the items in a  set often also consumed others items in the set, albeit not necessarily at the same time.
 HLTA can also determine the users who showed the tendency to consume the items in a co-consumption set. Those users make up a taste group.
To make recommendation for a user, we consider her memberships in the taste groups and past behaviors of those groups.
	We call this method {\em Conformative Filtering (CoF)} because a user is expected to conform to the behaviors of the groups she belongs to.
	
The main contributions of this paper include:
 \begin{enumerate}
 \item Proposing an intuitively appealing strategy, namely group UAN, to deal with the lack of negative examples; 
  \item Proposing a novel framework for OCCF, i.e., CoF, that is based on this assumption;
	\item Using HLTA, an algorithm proposed for text analysis, to solve a fundamental problem in collaborative filtering;

 \end{enumerate}
 \noindent The empirical results show that CoF significantly outperforms the state-of-the-art OCCF recommenders in predicting the items that users want to consume in the future.
In addition, the latent factors in CoF are more interpretable than those in matrix factorization methods.

	\section{Related Work} \label{related}
	
In the model-based approach to collaborative filtering, the goal is to find a feature vector
$\bf{f}_u$ for each user $u$ and a feature vector $\bf{f}_i$ for each item $i$, and predict the rating  of user $u$ for item $i$ using the inner products of the two vectors, i.e., $\hat{r}_{ui}=<\bf{f}_u, \bf{f}_i>$.
The dimension of $\bf{f}_u$ and $\bf{f}_i$ is usually much smaller than the number of users and the number of items.

Let $\mathcal{C}$ be the set all {\em consumption pairs}, i.e.,
user-item pairs $(u, i)$ such that $u$ consumed $i$ before.
The complement $\mathcal{U}$  of $\mathcal{C}$ consists of {\em non-consumption pairs}. In the case of explicit feedback data, we have a rating $r_{ui}$ for each pair $(u, i) \in \mathcal{C}$. It is the rating for item $i$ given by user $u$ and its possible values are usually  the integers between 1 and 5. The feature vectors can obtained by minimizing the following
loss function:
\begin{eqnarray*}
\sum_{(u, i) \in \mathcal{C}} (r_{ui} - \hr_{ui})^2 + \mbox{regularization terms}.
\label{eq.mf}
\end{eqnarray*}
\noindent In the literature, this is known as the matrix factorization (MF) method \cite{koren2009matrix}
 because $[\bf{f}_u]^{\top} [\bf{f}_i]$ is an approximate low-rank factorization of the user-item matrix $[r_{ui}]$.

 For implicit feedback data, researchers usually set $r_{ui}=1$ for consumption pairs $(u, i) \in \mathcal{C}$. There is no information about $r_{uj}$ for non-consumption pairs $(u, j) \in \mathcal{U}$. In this case, minimizing equation (\ref{eq.mf}) would lead to non-sensible solutions. Several methods have been proposed to solve the problem. We briefly review them below. Regularization terms and constraints are ignored for simplicity.

The sparse linear method (SLIM) \cite{ning2011slim}   makes the individual UAN assumption and
sets $r_{uj}=0$ for all $(u, j) \in \mathcal{U}$. It minimizes:
\begin{eqnarray*}
\sum_{(u, i) \in \mathcal{C}} (1 - \hr_{ui})^2 +
\sum_{(u, j) \in \mathcal{U}} (0 - \hr_{uj})^2 + \mbox{regularization terms}.
\label{eq.slim}
\end{eqnarray*}
\noindent  In addition, it lets $\bf{f}_u$ be the binary vector over items that represents past consumptions of user $u$, and it only finds $\bf{f}_i$.

The weighted regularized MF
(WRMF)\cite{Hu:2008:CFI:1510528.1511352,Pan:2008:OCF:1510528.1511402} algorithm makes the  UAN-with-low-confidence assumption and minimizes:
\begin{eqnarray*}
\sum_{(u, i) \in \mathcal{C}}(1 - \hr_{ui})^2 +
\sum_{(u, j) \in \mathcal{U}} c_{uj}(0 - \hr_{uj})^2+ \mbox{regularization terms},
\label{eq.WRFM}
\end{eqnarray*}
\noindent where $0 \leq c_{uj} \leq 1$ for all $(u, j) \in \mathcal{U}$. The values of the weights $c_{uj}$ indicate the confidence in treating the non-consumption pairs as negative examples.

The negative sampling method \cite{Pan:2008:OCF:1510528.1511402} makes the  UAN-with-chance assumption and minimizes:
\begin{eqnarray*}
\sum_{(u, i) \in \mathcal{C}} (1 - \hr_{ui})^2 +
\sum_{(u, j) \in \mathcal{U}'} (0 - \hr_{uj})^2+ \mbox{regularization terms},
\label{eq.negativeSampling}
\end{eqnarray*}
\noindent where  $\mathcal{U}'$ is a randomly sampled subset of $\mathcal{U}$.

The overlapping co-cluster recommendation (Ocular) algorithm \cite{heckel2017scalable} minimizes:
\begin{eqnarray*}
- \sum_{(u, i) \in \mathcal{C}}  \log (|1- e^{-\hr_{ui}}|)-
\sum_{(u, j) \in \mathcal{U}} \log (|0- e^{-\hr_{uj}}|)+  \mbox{regularization terms}.
\label{eq.ocular}
\end{eqnarray*}
\noindent This loss functions gives
large penalty if $\hr_{ui}$ is close to 0 for consumption pairs $(u, i)$ and small penalty if $\hr_{uj}$  is close to 1 for non-consumption pairs $(u, j)$. There is stronger ``force'' pushing $\hr_{ui}$ toward 1 and weaker ``force'' pushing $\hr_{uj}$ toward 0.
 So, ocular is implicitly making the UAN-with-low-confidence assumption.

The Bayesian personalized ranking MF (BPRMF) \cite{Rendle:2009:BBP:1795114.1795167} algorithm makes the  CPU assumption and minimizes:
\begin{eqnarray*}
\sum_u \sum_{i \in \mathcal{C}_u} \sum_{j \in \mathcal{U}_u} - \log \sigma(\hr_{ui} - \hr_{uj})+ \mbox{regularization terms},
\label{eq.BPR}
\end{eqnarray*}
\noindent
\noindent where $\mathcal{C}_u =\{i | (u, i) \in \mathcal{C}\}$, $\mathcal{U}_u =\{j | (u, j) \in \mathcal{U}\}$, and $\sigma$ is the sigmoid function. The penalty for a user $u$ is small if the predicted scores $\hr_{ui}$ for the consumed items are large relative to the predicted scores $r_{uj}$ for the unconsumed items.

Various extensions of the aforementioned methods  have been proposed. SLIM has been extended by \cite{cheng2014lorslim,christakopoulou2014hoslim,levy2013efficient}, WRMF has been extended by \cite{He:2016:FMF:2911451.2911489,Pilaszy:2010:FAM:1864708.1864726,Wang2016}, and BPRMF has been extended by \cite{hong2012learning,pan2013cofiset,pan2013gbpr,takacs2012alternating}.

\begin{figure*}[t]
		\centering
		\includegraphics[width=5in,height=2in]{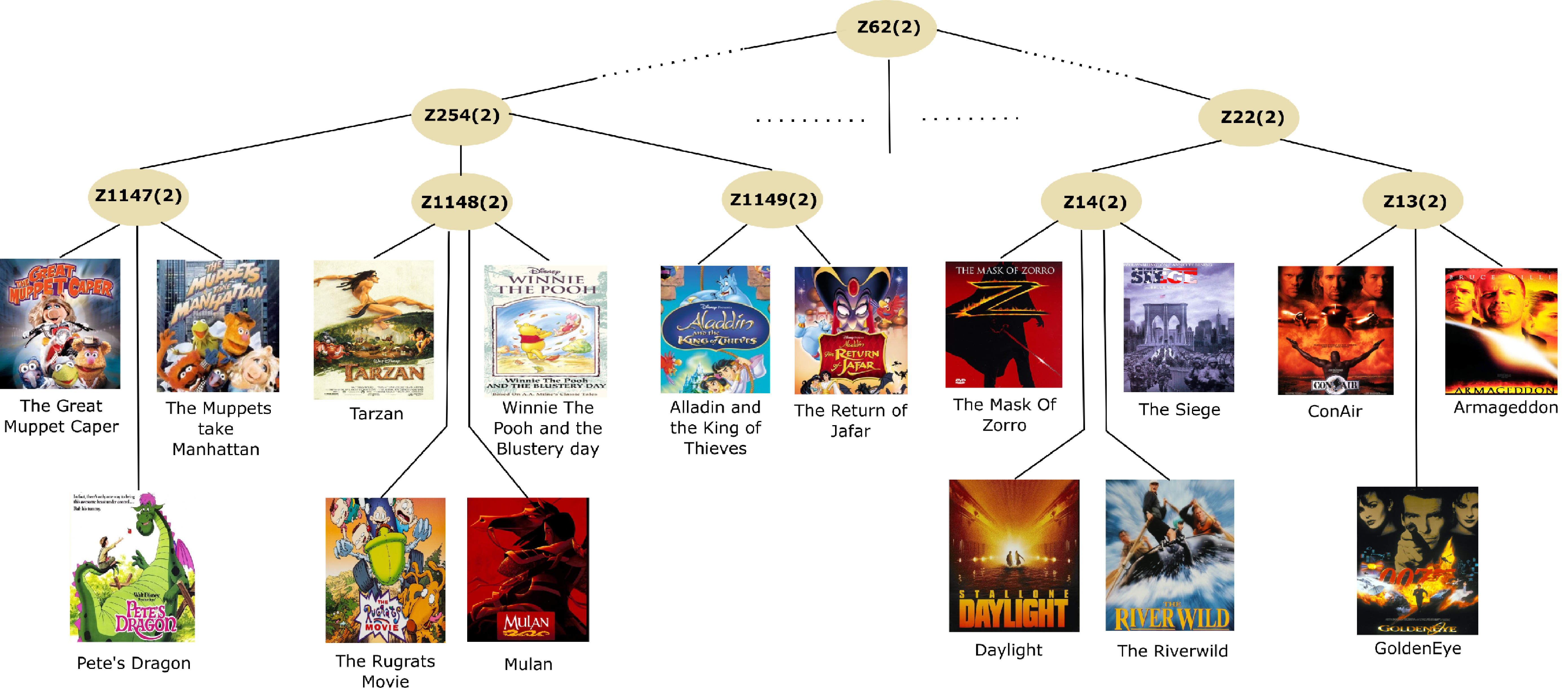}
		\vspace{-4mm}
		\caption{A part of the hierarchical latent tree model learned from Movielens dataset.   The level-1 latent variables reveal co-consumption of items by users and identify user tastes for various subsets of items. Latent variables at higher levels reveal co-occurrence of the tastes at the level below and identify more broad tastes.}
		
		\label{fig:Movielens1M}
	\end{figure*}

Clustering algorithms have also been applied to CF. The k-means algorithm and hierarchical clustering have been used to group either users or items to reduce the time complexity of CF methods such as user-kNN and item-kNN \cite{Aggarwal:2016:RST:2931100,Koren2015}.  Co-clustering has been used to identify user-item co-clusters so as to model user group heterogeneity \cite{heckel2017scalable,Wu:2016:CIC:2835776.2835836,Xu:2012:EIC:2187836.2187840}. In \cite{Wu:2016:CIC:2835776.2835836,Xu:2012:EIC:2187836.2187840}, the authors find multiple sub-matrices (co-clusters) of the user-item matrix, apply another CF method on each sub-matrix, and then aggregate the results. In ocular \cite{heckel2017scalable}, the authors first obtain feature vectors $f_u$ and $f_i$, and then use those vectors to produce user-item co-clusters for the sake of interpretability of the results. The notion of user groups is used in an extension of BPRMF called GBPR \cite{pan2013gbpr}. In GBPR a group of users is formed for each consumption pair $(u, i)$, and it consists of  a few randomly selected other users who also consumed the item $i$ before. In CoF a user taste group is determined based on a set of items that tend to be co-consumed, and it is typically quite large. However, none of the aforementioned methods use the (user, item, or user-item) clusters obtained to deal with the lack of negative preference in implicit feedback data.

\section{User taste Group Detection using HLTA} \label{TGD}

	\begin{table*}[t]
		\caption{User clusters identified by the latent variables $Z_{13}$ and $Z_{1147}$. High percentages of the users in the cluster $Z_{13}=s_1$ have watched the three movies
{\tt Armageddon}, {\tt Golden Eye} and {\tt Con Air}. Hence, the cluster is regarded as a user group with a taste for the three movies. Similarly, the cluster $Z_{1147}=s_1$
is regarded as a user group with a taste for the three movies {\tt The Great Muppet Caper}, {\tt Pete's Dragon}  and {\tt The Muppets take Manhattan}.
	}
	\vspace{-3mm}
		\label{table.Z13}
		\begin{minipage}{.48\linewidth}
			
			\centering
			{\small
				\begin{tabular}{lcc}
					\hline
					& $\scriptstyle Z_{13}=s_1$ & $\scriptstyle Z_{13}=s_0$\\
					
					{\tt Action-Adventure-Thriller}& (0.21) &  (0.79) \\
					\hline
					$\scriptstyle {\tt Armageddon} $   & 0.610& 0.055 \\
					
					$\scriptstyle {\tt Golden\_Eye}$  & 0.588 & 0.013 \\				 
					
					$\scriptstyle {\tt Con\_Air} $  & 0.635  & 0.014 \\
					
					\hline
				\end{tabular}
			}
		\end{minipage}%
		\hspace{10px}
		\begin{minipage}{.48\linewidth}
			\centering
			
			{\small
				\begin{tabular}{lcc}

					\hline
					& $\scriptstyle Z_{1147}=s_1$ & $\scriptstyle Z_{1147}=s_0$\\
					
					{\tt Children-Comedy} & (0.09) &  (0.91) \\
					\hline
					$\scriptstyle {\tt Great\_Muppet\_Caper\_The} $   &0.456 &0.009  \\
					
					$\scriptstyle {\tt Petes\_Dragon}$  & 0.450 & 0.004 \\
					
					$\scriptstyle {\tt Muppets\_Take\_Manhattan\_The}$  & 0.457  & 0.005 \\

					\hline
				\end{tabular}
			}
		\end{minipage}
	\end{table*}

When applied to implicit feedback data, HLTA\footnote{https://github.com/kmpoon/hlta.} learns models such as the one shown in Figure \ref{fig:Movielens1M}, which was obtained from the Movielens dataset.
Movielens is an explicit feedback dataset. It was turned into an implicit feedback dataset by ignoring the item ratings\footnote{Movielens is used for illustration since movies genres are easier to interpret.}.

The model is a tree-structured Bayesian network, where there is a layer
of observed variables at the bottom, and multiple layers
of latent variables on top. It is called a {\em hierarchical latent tree (HLTM)} model \cite{chen2017latent,liu2014hierarchical}.
The model parameters include a marginal distribution for the root\footnote{When there are multiple latent variables at the top level, arbitrarily
pick one of them as the root.} and a conditional distribution for each of the other nodes given its parent. The product of the distributions defines a joint distribution over all the variables.

In this paper, all the variables are assumed to be binary. The observed variables indicate whether the items were consumed by a user. For example, the value of the variable
{\tt Mulan} for a user is 1 if she watched the movie before, and 0 otherwise. Note that here the value 0 means non-consumption, not disinterest.

The latent variables are introduced during data analysis to explain co-consumption patterns detected in data.
	For example, the fact that the variables {\tt Armageddon}, {\tt Golden Eye} and {\tt Con Air} are grouped under $Z_{13}$ indicates that the three movies  tend to be  co-consumed, in the sense users who watched one of them often also watched the other two. The pattern is explained by assuming that there is a taste, denoted by $Z_{13}$, such that users with the taste tend to watch the movies and users without it do not tend to watch the movies.
Similarly,	 $Z_{14}$ explains the co-consumption of {\tt The Seige}, {\tt Mask of Zorro}, {\tt Daylight} and {\tt The River Wild}. $Z_{22}$ indicates that the patterns represented by $Z_{13}$ and $Z_{14}$ tend to co-occur.

HLTMs  are a generalization of latent class models (LCMs) \cite{knott1999latent}, which is a type of finite mixture models for discrete data. In a finite mixture model, there is one latent variable and it is used to partition objects into soft clusters. Similarly, in an HLTM, each latent variable  partitions all the users into two clusters.
Since there are multiple latent variables, multiple partitions are obtained. In this sense, HLTMs are a tool for {\em multidimensional clustering} \cite{chen2012model,liu2015greedy,Zhang:2004:HLC:1005332.1016782}.

Information about the partition given by
$Z_{13}$  is shown in Table \ref{table.Z13}. The first cluster $Z_{13}=s_1$ consists of 21\% of the users. High percentages of the users in the cluster have watched
the three movies {\tt Armageddon}, {\tt Golden Eye} and {\tt Con Air}. So, they have a taste for them. In contrast, few users in the second cluster $Z_{13}=s_0$ have watched these movies and hence they do not possess the taste.

Similarly, $Z_{14}$ identifies another group of users with a taste for the movies  {\tt The Seige}, {\tt Mask of Zorro}, {\tt Daylight} and {\tt The River Wild}. $Z_{14}$ and $Z_{13}$ are grouped under $Z_{22}$ in the model structure, which indicates that the two tastes tend to be co-possessed, and $Z_{22}$ identifies the users who tend to have both tastes.

\section{Conformative Filtering} \label{CoF}
Suppose we have learned an HLTM $m$ from an implicit feedback
dataset and suppose there are $K$ latent variables on the
$l$-th level of the model, each with two states $s_0$ and $s_1$. Denote
the latent variables as $Z_{l1}, \ldots, Z_{lK}$. They give us $K$
 user taste groups $Z_{l1} = s_1, \ldots, Z_{lK} = s_1$, which
will sometimes be denoted as $G_1, \ldots , G_K$ for simplicity. In
this section, we explain how these user taste groups can be used for
item recommendation.

	\subsection{User Group Characterization}
	A natural way to characterize the preferences of a user group for items is to aggregate past behaviors of the group members. The issue is somewhat complex for us because our user groups are soft clusters. Let $\mathbb{I}(i|u, {\mathcal D})$ be the indicator function which takes value 1 if user $u$ has consumed item $i$ before, and 0 otherwise. We determine the preference of a taste group $G_k$ (i.e., $Z_{lk}=s_1$) for an item $i$ using the {\em relative  frequency that the item was consumed by users in the group}, i.e.:
	
	\begin{eqnarray}
		\phi(i|G_k, {\mathcal D}) = \frac{\sum_{u} \mathbb{I}(i|u, {\mathcal D})
			P(G_k|u, m)}{\sum_{u} P(G_k|u, m)},
		\label{eq.group}
	\end{eqnarray}
	\noindent where $P(G_k|u, m)$ is the probability of user $u$ belonging to
	group $G_k$, and the summations are over all the users who consumed item $i$ before.

	Note that $\phi(i|G_k, {\mathcal D})=0$ if no users in $G_k$ have consumed the item $i$ before.
	In other words, we assume that a group is not interested in an item if none of the group members have consumed the item before.

There is an important remark to make.	The reason we determine the preferences of a user group $G_k$ is that we want to predict future behavior of the group. As such, we might want to base the prediction on recent behaviors of the group members instead of their entire consumption histories. For example, we might want to choose to use
a subset  ${\mathcal D}_H$ of the data that consists of only the latest $H$ consumptions for each user. We will empirically investigate this strategy and will show that the choice of $H$  has an impact on the quality of item recommendations.
	
	\subsection{Item Recommendation}

Having characterized the user taste groups, we now give feature vectors for items and users. We characterize item $i$ using a vector where the $k$-th component is the relative frequency that it was consumed by members of group $G_k$, i.e.,
\begin{eqnarray}
\mathbf{f}_i = (\phi(i|G_1, {\mathcal D}_H), \ldots, \phi(i|G_K, {\mathcal D}_H)).
\label{eq.fi}
\end{eqnarray}

\noindent Note that ${\mathcal D}_H$ is used instead of ${\mathcal D}$, which means that the latent representation is obtained from the  $H$ most recent consumptions of users. 

We characterize	 user $u$ using a vector where the $k$-th component is the probability that user $u$ belongs to the group $G_k$, i.e.,
	\begin{eqnarray}
		\mathbf{f}_u= (P(G_1|u, m), \ldots, P(G_K|u, m)).
		\label{eq.fu}
	\end{eqnarray}
	
	The latent representations require the computation the posterior probabilities $P(G_k|u, m)=P(Z_{lk}=s_1|u, m)$  for $k=1, \ldots, K$. Because $m$ is a tree-structured model, all the posterior probabilities  can be computed by propagating messages over the tree twice \cite{Pearl:1988:PRI:52121}. It takes time linear in the number of variables in the model, and hence linear in the number of items.
	
We use the inner product of the two vectors $\bf{f}_i$ and $\bf{f}_u$ as the	 predicted score $\hat{r}_{ui}$ for the user-item pair $(u, i)$, i.e.,
	\begin{eqnarray}
		\hat{r}_{ui} = \sum_{k=1}^K \phi(i|G_k,{\mathcal D}_H) P(G_k|u, m).
		\label{eq.score}
	\end{eqnarray}

\noindent To make recommendations for a user $u$, we  sort all the items $i$ in descending order of the predicted scores $\hat{r}_{ui}$, and recommend  the items with the highest scores.

	\subsection{Discussions}
Matrix factorization (MF) is often used in collaborative filtering to map items and users to
feature vectors in the same Euclidean space. The components of the vectors are called {\em latent factors}, which are not be confused with latent variables in latent tree models.

 CoF   differs fundamentally from  MF. The latent factors in MF
 are obtained by factorizing the user-item matrix and they are not interpretable.
 In contrast, the latent factors in CoF are characteristics of user taste groups and they have clear semantics.

    CoF naturally incorporates the group UAN assumption, which is the most reasonable assumption to deal with the lack of negative examples to date. In contrast,
 MF has only been extended to incorporate the individual UAN assumption and its variants. It cannot  incorporate the group UAN assumption because there is no notion of user groups.

 In addition, CoF has a desirable characteristic that is not shared by MF. It considers the entire consumption histories of the users when grouping  them and uses only recent  user behaviors when predicting what the groups would like to consume in the future. Consumption behaviors long ago are useful when identifying similar users. However, they might not be very useful when predicting what the users would like to consume in the future. Actually, they might be misleading.

	\section{Experiments}\label{EXP}
	We performed experiments on two real-world datasets to compare CoF with five baselines using two evaluation metrics.

\subsection{Datasets}

The datasets used in our experiments are
Movielens20M\footnote{\scriptsize \url{https://grouplens.org/datasets/movielens/20m/}} and
{Ta-feng}\footnote{\scriptsize
\url{www.bigdatalab.ac.cn/benchmark/bm/dd?data=Ta-Feng}}.
Movielens20M contains ratings given by users to the movies they watched.
It is an explicit feedback dataset and was converted to implicit feedback data by keeping all the rating events and  ignoring the rating values. Ta-feng contains purchase events at a supermarket, where each event is a customer-item pair and a checking-out action by a customer involves multiple events.

Statistics about the datasets are as follows:
\begin{center}
{\scriptsize
				\begin{tabular}{lccc}

                    \hline
					\textbf{Movielens20M}& \textbf{Users}  &\textbf{Items}  & \textbf{Sparsity} \\
					\hline
					train&118,526&	15,046&	99.047\% \\
					validation& 22,684&	14,888&	99.112\%  \\
					test & 25,561	&25,843&	99.546\% \\

					\hline
					\textbf{Ta-feng}& &&  \\
					\hline
					train& 27,574 &22,226  & 99.907\% \\
					validation&12,261  &15,206  & 99.934\% \\
					test& 13,191 &14,561  &99.936\%  \\

					\hline
				\end{tabular}
			}
\end{center}

Each dataset is comprised  of \textit{(user, item, time-stamp)} tuples.
Following \cite{Wu:2017:RRN:3018661.3018689}, we split each dataset into training, validation and test sets by time. This is so that  all the training instances came before all the testing instances, which matches real-world scenarios better than splits that do not consider time. We tested on several splits and the results were similar. In the following, we will only report the results on the split with 70\% of the data for training, 15\% for  validation and 15\% for test.

\subsection{Baselines}

In the Related Work section, we discussed five representative OCCF methods, namely WRMF, BPRMF, ocular (co-clustering), GBPR (group based) and SLIM (model based neighborhood method).
They were all included in our experiments.\footnote{SLIM failed to finish a single run in one week on the Movielens20M dataset during validation, therefore its performance is not reported on this dataset.}
 The implementation of the original authors was used for ocular, and  the LibRec implementations\footnote{\url{https://www.librec.net/index.html}} and the MyMediaLite implementations \cite{Gantner:2011:MFR:2043932.2043989} were used for all other baselines. 
	
	All the algorithms require some input parameters. For the baselines, we tuned their parameters on the validation set through grid search, as is commonly done in the literature. The best parameters were chosen based on recall@$R$. The details of the parameters chosen can be found in Table \ref{tab.param}.  The key parameters are: $F$ --- the number of latent factors; $\lambda$, $\beta/2$ --- weights for regularization terms; $k$ --- the size of neighborhood; $|G|$ --- group size; $\rho$ --- tuning parameter. We refer the reader to the original papers for the meanings of other parameters.

	\begin{table}[]
	\centering
	\caption{Parameters selected by validation.}
		\label{tab.param}
	\scriptsize
	\begin{tabular}{lllllll}
		\hline
		&\textbf{Movielens20M}  &  &  \textbf{Ta-feng}     &&                            \\
		\hline
		CoF         & $l$=1,$H$=5           && $l$=1,$H$=40  \\
		WRMF        & $F$=40,$\lambda$=$10^{-2}$       && $F$=10,$\lambda$=10   \\
		BPRMF       & $F$=80,$\lambda$=$10^{-2}$  && $F$=80,$\lambda$=$10^{-4}$  \\
		Ocular      & $F$=120,$\lambda$=80        && $F$=60,$\lambda$=120 \\
		SLIM &N.A. && $k$=500,$\lambda$=$10^{-2}$,$\beta/2$=10&\\
		GBPR &{$F$=160,$\lambda=10^{-4}$, $|G|=10$,$\rho = 0.2$}    &&$F$=20,$\lambda$=0.1, $|G|=4$,$\rho = 0.4$ \\

		\hline
	\end{tabular}
	\end{table}	
For CoF, we searched for the value of  $H$ over the set $\{2,3,4,5,10,20, \dots,100\}$, and for $l$ we considered all levels of the hierarchical model obtained by HLTA.
 The number $K$ of latent variables on a level of the model was automatically determined by HLTA.

\subsection{Evaluation Metrics}

Two standard evaluation metrics are used in our experiments, namely
 recall@$R$, and area under the curve (AUC).
The evaluation metrics are briefly described below:
\begin{itemize}
	\item Recall@$R$: It is the fraction, among all items consumed by a user in the test set, of those that are placed at the top $R$ positions in a recommended list. Formally, recall is defined as: $\frac{TP}{TP+FN}$, where $TP$ denotes true positive and $FN$ denotes false negative.
	
	\item AUC : Is the area under the Receiver Operating Characteristic curve.  It is the probability that  two items randomly picked from the recommendation list are in the correct order, i.e., the first is a consumed item and the second is an unconsumed item. 
\end{itemize}
	\subsection{Results}
	
\begin{figure}[]
	\centering
	\setlength{\fboxrule}{0pt}
		\fbox{

		\includegraphics[scale=0.3]{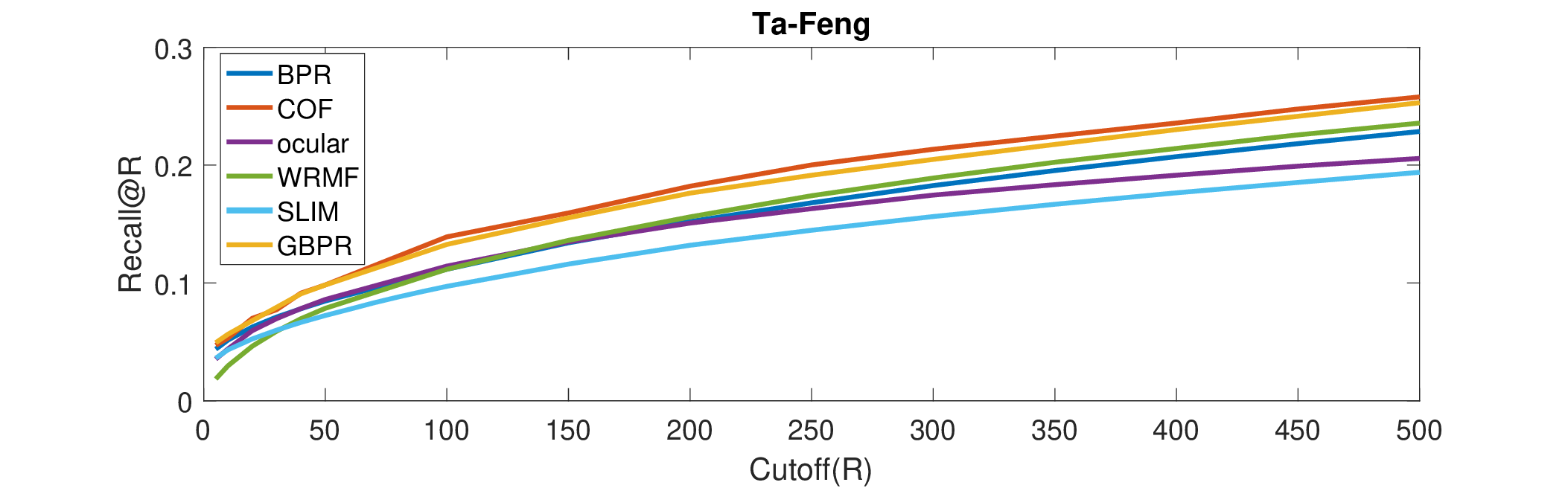}
		}

		\setlength{\fboxrule}{0pt}
		\fbox{

		\includegraphics[scale=0.3]{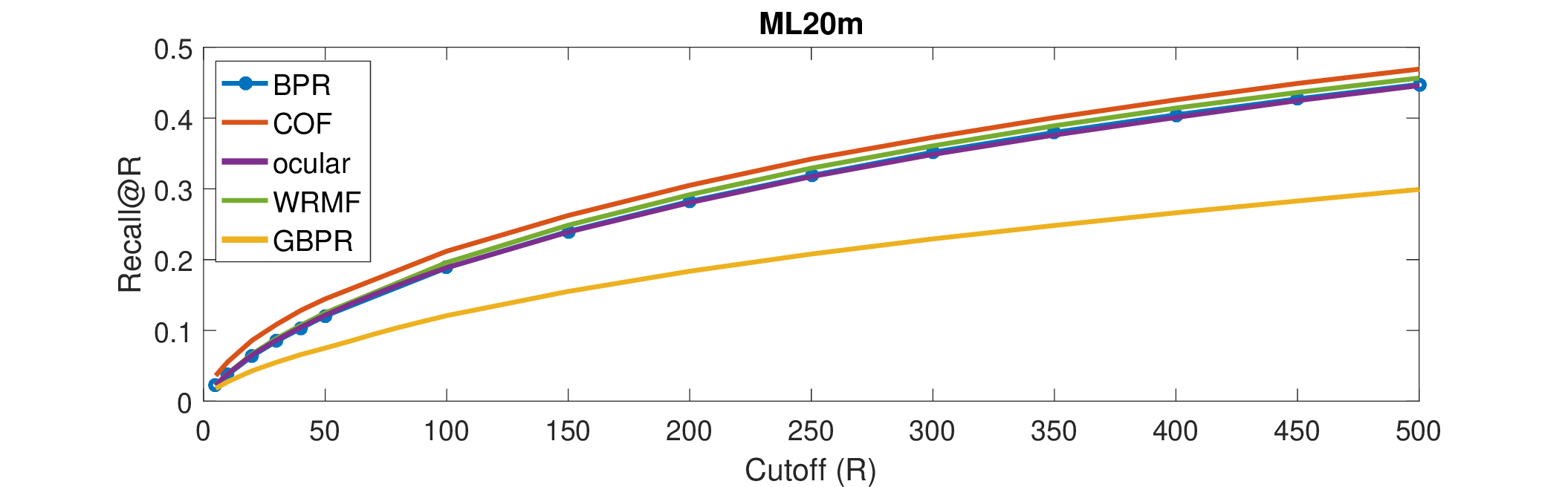}}
		\vspace{-3mm}
		\caption{The recall@R performance of different recommenders on the ta-feng and Movielens20M dataset. CoF exhibits the best performance on both datasets due to better representation of individual tastes.}
		\label{figure.results}
	\end{figure}

    The recall@$R$ results, at each cutoff position $R$, are shown in Figure \ref{figure.results}. We see that CoF achieved the best performance on both datasets. Since CoF attempts to model each taste of a user individually, a higher recall suggests that these tastes are catered for while making recommendations. The improvements are comparatively larger on the Movielens20M dataset. This indicates that when the data is relatively less sparse, CoF is able to extract meaningful information much more effectively than other methods. 

Ocular performed competitively with BPR and WRMF in terms of recall@R. However, WRMF and BPRMF performed better at larger cutoff values. Interestingly, we found that the performance of GBPR was better than all baselines on the ta-feng dataset but was unimpressive on Movielens20M despite extensive parameter tuning. A possible reason for this could be the tendency of GBPR to focus on popular items\footnote{During our experiments we found that GBPR has low global diversity. These results are not reported in the interest of space.}. Ta-feng is grocery dataset and certain common items are in every customer's basket (e.g. bread), focusing on these may lead to a higher recall. On the other hand, the movie domain is comparatively more personalized and focusing on popular items might not cater to different tastes of a user.

Table \ref{tab:AUC} shows the performance in terms of AUC. CoF outperforms the baselines on both datasets. This suggests that CoF is able to identify the ``true'' negatives and it puts them lower than the items of interest in the ranked list. We see that BPRMF performs the second best w.r.t. AUC on both datasets. This is expected since BPRMF optimizes for AUC. Moreover, we note that WRMF performed better than BPRMF in terms of recall@$R$, however, it's performance in terms of AUC is lower. This provides further evidence that the score of the top-$R$ metrics and those which evaluate over the whole list might not correlate and depending on the target of the recommendation an appropriate metric should be chosen.
\begin{table}[]
\centering
\caption{The AUC for each recommender is shown. CoF outperforms other methods. }
\vspace{-3mm}
\label{tab:AUC}
\begin{tabular}{lllllll}
\hline
&\textbf{BPRMF}   & \textbf{WRMF}    & \textbf{CoF}     & \textbf{Ocular}  &\textbf{ SLIM} & \textbf{GBPR}           \\
\hline
Ta-feng & 0.74977 & 0.71316 & \textbf{0.7793}  & 0.63653 & 0.68321&  0.71117\\

ML-20M  & 0.87289 & 0.85258 & \textbf{0.88816} & 0.84879  & N.A.& 0.80367 \\
\hline      
\end{tabular}
\end{table}
SLIM gave the lowest performance over all metrics in our experiments. It is worth noting that our experimental setup (splitting the data by time) and the metrics used differ from the experimental conditions under which SLIM is normally evaluated.
	
	\subsection{Impact of Parameters}

	\begin{figure}[]
	\centering
	\setlength{\fboxrule}{0pt}
		\fbox{\includegraphics[width=2.2in,height=1.5in]{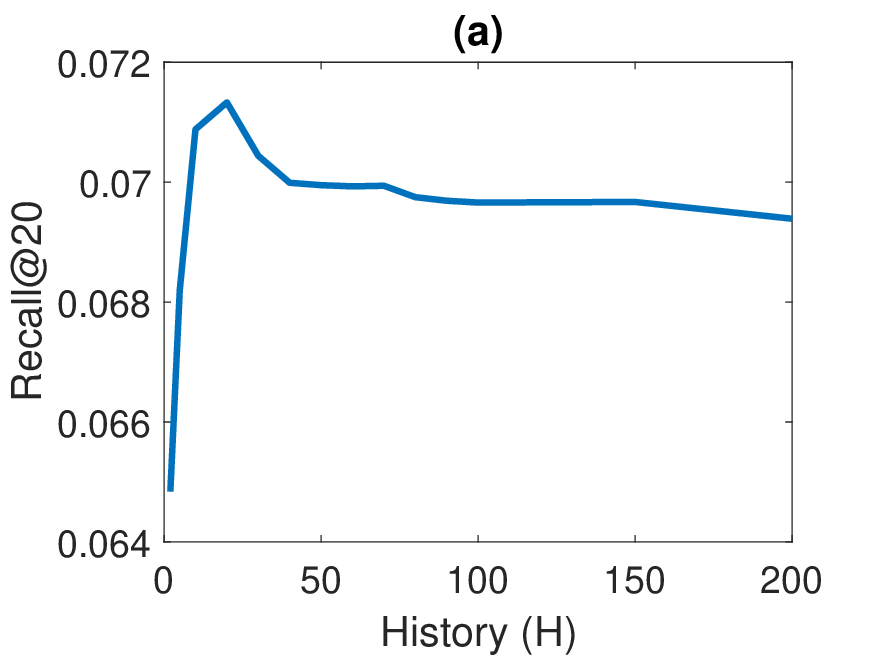}
		
		\includegraphics[width=2.2in, height=1.5in]{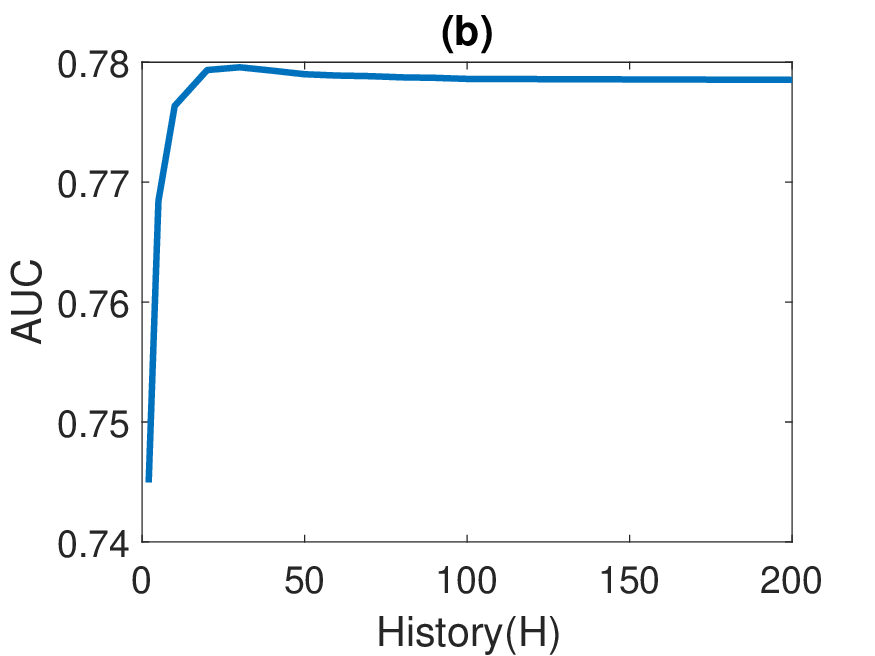}}
		\vspace{-3mm}
		\caption{Impact of parameters on the performance of CoF on ta-feng dataset.}
		\label{fig:H}
	\end{figure}
	
CoF begins by running HLTA to learn a hierarchical model and then uses
the model for item recommendation. There are two parameters. The first parameter $l$ determines which level of the hierarchy to use. The larger the $l$, the fewer the number of user taste groups. The second parameter $H$ determines the amount of consumption history to use when characterizing user groups.
 Although both parameters are selected via validation, it would be interesting to gain some insights about how they impact the performance of CoF.
	
	Figure \ref{fig:H}(a) shows the recall@$20$ scores on ta-feng as a function of $H$ when $l=1$. We see that the curve first increases with $H$ and then decreases with $H$. It reaches the maximum value when $H=20$. The reason is that, when $H$ is too small,  the data used for user taste group characterization contain too little information. When $H$ is too large, on the other hand,  too much history is included and the data do not reflect the current interests of the user groups.
	
	Figure \ref{fig:H}(b) shows the AUC scores on ta-feng as a function of $H$ when $l=1$. We observe a similar trend, but the impacts of $H$ are not as pronounced on AUC. CoF is more or less robust to the choice of $l$ and the performance is almost the same regardless of the level chosen. The results are not shown for brevity. This was somewhat unexpected as when $l$ increases the taste become more general and one would expect the performance to deteriorate\footnote{We did observe slight deterioration but the magnitude was too small to draw conclusions from.}.

	\section{Conclusion}
A novel method called CoF is proposed for collaborative filtering with implicit feedback data. It deals with the lack of negative examples which does not perform negative sampling, rather it uses the group
UAN assumption, which is more reasonable than assumptions made by previous works.  Extensive experiments were performed to compare CoF with a variety of baselines on two real-world datasets. CoF achieved the best performance over both recall@$R$ and AUC signifying that various taste of an individual is captured and the true negatives are placed at the bottom of the ranked list.

\section*{Acknowledgment}
Research on this article was supported by Hong Kong Research Grants Council under grant 16202118.

\bibliographystyle{splncs04}
\bibliography{COF}

\end{document}